\documentclass[12pt]{iopart}
\usepackage[dvips]{graphicx}
\newcommand{\myext}{ps}

\newcommand{\JCIS}{{\it J.~Coll.~Int.~Sci.~}}
\newcommand{\EPL}{{\it Europhys.~Lett.~}}

\newcommand{\kt}{k_{\rm B} T}

\newcommand{\rv}{{\bf r}}
\newcommand{\Rv}{{\bf R}}
\newcommand{\xv}{{\bf x}}

\newcommand{\vcc}{v_{\colloid\colloid}}
\newcommand{\vcp}{v_{\colloid\polymer}}
\newcommand{\vpp}{v_{\polymer\polymer}}

\newcommand{\n}[2]{n_{#1}^{#2}}
\newcommand{\nv}[2]{{\bf n}_{{\rm v}#1}^{#2}}

\newcommand{\w}[2]{w_{#1}^{#2}}
\newcommand{\wv}[2]{{\bf w}_{{{\rm v}#1}}^{#2}}

\newcommand{\wm}[2]{{\bf \hat{w}}_{{{\rm m}#1}}^{#2}}
\newcommand{\mymat}[1]{{\hat{\mathbf{#1}}}}

\newcommand{\nmtl}[2]{{\mymat{n}_{{\rm m}#1}}^{#2}}

\newcommand{\colloid}{C}
\newcommand{\polymer}{P}
\newcommand{\speciesindex}[2]{#1_{#2}}

\newcommand{\etai}{\speciesindex{\eta}{i}}
\newcommand{\etac}{\speciesindex{\eta}{\colloid}}
\newcommand{\etap}{\speciesindex{\eta}{\polymer}}
\newcommand{\ec}{\etac}

\newcommand{\rhoc}{\speciesindex{\rho}{\colloid}}
\newcommand{\rhop}{\speciesindex{\rho}{\polymer}}

\newcommand{\Rc}{\speciesindex{R}{\colloid}}
\newcommand{\Rp}{\speciesindex{R}{\polymer}}

\newcommand{\Rg}{\speciesindex{R}{\rm g}}

\newcommand{\myphi}[1]{{\speciesindex{\varphi}{#1}}}

\newcommand{\fch}{f_{\rm chain}}
\newcommand{\fumindex}{\nu}

\newcommand{\phihs}{\phi_{\rm hs}}

\begin{document}

\title{Colloid-induced polymer compression}

\author{Alan R Denton\dag\ and Matthias Schmidt\ddag
\footnote[3]{Authors contributed equally}}
\address{\dag\ Department of Physics,
North Dakota State University, Fargo, ND 58105-5566, USA \\
alan.denton@ndsu.nodak.edu}
\address{\ddag\ Institut f{\"u}r Theoretische Physik II, 
Heinrich-Heine-Universit{\"a}t D{\"u}sseldorf, 
Universit{\"a}tsstra{\ss}e 1, D-40225 D{\"u}sseldorf, Germany \\
mschmidt@thphy.uni-duesseldorf.de}

\date{30 May 2002}
\begin{abstract}
We consider a model mixture of hard colloidal spheres 
and non-adsorbing polymer chains in a theta solvent.
The polymer component is modelled as a polydisperse mixture
of effective spheres, mutually noninteracting but excluded 
from the colloids, with radii that are free to adjust to 
allow for colloid-induced compression.
We investigate the bulk fluid demixing behaviour of this model system 
using a geometry-based density-functional theory that includes 
the polymer size polydispersity and configurational free energy, 
obtained from the exact radius-of-gyration distribution for 
an ideal (random-walk) chain.  Free energies are computed by 
minimizing the free energy functional with respect to the 
polymer size distribution.
With increasing colloid concentration and polymer-to-colloid 
size ratio, colloidal confinement is found to 
increasingly compress the polymers.
Correspondingly, the demixing fluid binodal shifts, compared to 
the incompressible-polymer binodal, to higher polymer densities 
on the colloid-rich branch, stabilizing the mixed phase. 
\end{abstract}

\maketitle
\newpage

\section{Introduction}
\label{SECintroduction}
The physical properties of soft matter systems are often 
influenced by size polydispersity 
on mesoscopic length scales~\cite{daoud-williams,sollich02}.
In colloidal suspensions, for example, variations in particle size 
can profoundly affect fluid-solid transitions~\cite{bolhuis-kofke}.
In polymer solutions, chain length (degree of polymerization)
often has a broad distribution.  In practice, distributions in 
colloidal particle radius and polymer chain length often can be 
narrowed by physical selection or special synthesis methods. 
However, even monodisperse polymer chains have a radius of gyration 
that fluctuates in response to the surrounding environment.
In this sense, polymers in solution are fundamentally polydisperse.

Polydispersity is especially relevant in mixtures of colloids 
and nonadsorbing (free) polymers, in which the polymer size sets
the range of effective depletion-induced interactions between 
colloidal particles. A popular basis for modelling colloid-polymer
(CP) mixtures is the Asakura-Oosawa (AO) model~\cite{AO}. 
The AO model neglects fluctuations in chain conformations by 
approximating the polymers as effective spheres of fixed radius
that are excluded from the hard-sphere (HS) colloids. 
The model further ignores polymer-polymer interactions by
treating the polymer spheres as freely interpenetrating, 
an approximation at best justified in a theta solvent.

Despite its simplicity, the AO model qualitatively captures 
thermodynamic phase behaviour observed in real CP mixtures -- 
in particular, bulk fluid demixing into colloid-rich and -poor 
phases~\cite{gast86,pusey91,lekkerkerker92,ilett95}.
To address quantitative discrepancies with experiment, 
several recent studies have gone further
to include polymer-polymer interactions -- in both good
solvents~\cite{warren95,schmidt-intpol02,bolhuis-louis02} and
poor~\cite{schmidt-poorsolvent02} -- and polydispersity
in polymer chain length~\cite{sear97,warren97,lee99}.
In all of these studies, however, the polymers are modelled
as effective spheres of fixed size (or fixed size distribution).

An issue that has not yet been widely explored is the influence of 
colloidal confinement on polymer conformations and implications 
for phase behaviour of CP mixtures. 
Polymers near surfaces and polymer-mediated interactions 
between surfaces have been examined by a variety of theoretical approaches, 
including scaling theory~\cite{degennes,joanny79,degennes81}, 
integral equation theory~\cite{yethiraj92,chatterjee98},
classical density-functional (DF) theory~\cite{bechinger99}, 
and field-theoretic methods~\cite{eisenriegler96}.
A far more challenging problem is the influence on polymers of
confinement to a fluctuating random porous medium created by 
colloidal particles in suspension.

Numerical studies of colloids mixed with segmented polymer chains 
give some insight into polymer conformations in CP mixtures.
Meijer and Frenkel~\cite{meijer-frenkel91} performed
Monte Carlo simulations of hard spheres mixed with ideal 
(noninteracting) lattice-polymers.
Dickman and Yethiraj~\cite{dickman-yethiraj94} simulated mixtures
of hard spheres and off-lattice polymer chains, modelled as 
freely-jointed ``pearl necklaces" of hard spheres. 
Simulations such as these have revealed the potential importance 
of fluctuations in polymer size and shape in CP mixtures.

Here we consider an intermediate model in which the polymers 
maintain spherical shape but can adjust their radius-of-gyration 
distribution to the bulk colloid concentration.
The constraint of fixed polymer size is thus relaxed by 
endowing the effective polymer spheres with an internal 
degree of freedom that can respond to colloidal confinement.
We study the phase behaviour of a model mixture of HS colloids 
and compressible polymers by means of a geometry-based DF theory.  
Confinement-induced adjustments in polymer size are permitted 
by including in the free energy functional the free energy of 
ideal polymers, derived from the exact radius-of-gyration distribution 
of a random-walk chain. 

The theory, when applied to bulk fluid phases, predicts both the 
demixing binodal and the polymer size distribution as a function 
of colloid and polymer concentrations.  
For sufficiently large polymer-to-colloid size ratio,
the model displays bulk fluid demixing into colloid-rich 
and -poor phases, qualitatively similar to the behaviour
of the AO model.  Upon demixing, however, the relatively 
unconfined polymers in the colloid-poor (vapour) phase coexist 
with compressed polymers in the colloid-rich (liquid) phase.
We find that the polymer radius of gyration can be significantly
reduced, by up to 20 percent or more, in the colloidal liquid phase.
Correspondingly, the fluid demixing binodal shifts relative to 
the incompressible-polymer binodal.

Next, in Sec.\ \ref{SECmodel}, we explicitly define the model system.
In Sec.\ \ref{SECtheory}, we construct an appropriate 
classical DF theory that incorporates the conformational 
free energy of the polymer chains.
Results for demixing phase diagrams and polymer size distributions 
are presented in Sec.\ \ref{SECresults}.  Finally, 
we conclude in Sec.\ \ref{SECconclusions}.

\section{Model}
\label{SECmodel}
The model we consider is a generalization of the AO model 
to the case of compressible polymers.
Explicitly, we consider a mixture of hard colloidal spheres 
(species $\colloid$), of monodisperse radius $\Rc$, and 
non-adsorbing, linear polymer chains (species $\polymer$), 
monodisperse in length, but polydisperse in radius of gyration, $\Rp$,
in a volume $V$ at a temperature $T$ near the theta temperature. 
The restriction to a theta solvent allows the polymer chains 
to be reasonably approximated as noninteracting random walks. 
The colloids interact via a hard-sphere pair potential: 
$\vcc(r)=\infty$, if $r<2\Rc$, zero otherwise, where $r$ is 
the centre-centre interparticle distance.
When interacting with colloids, each polymer is assumed to
behave as an effective hard sphere of radius equal to
its radius of gyration:
$\vcp(r)=\infty$, if $r<\Rc+\Rp$, zero otherwise.
Finally, interactions between polymers vanish for all distances: 
$\vpp(r)=0$.  

To fix the polymer chemical potential, it is convenient to imagine 
a reservoir of pure polymer solution with which the system can freely 
exchange polymer (but not colloid) through a semi-permeable membrane.
The reservoir also serves as a colloid-free reference state in which
the polymer assumes an ideal radius-of-gyration distribution. 
Bulk fluid states are specified by the mean colloid and polymer (reservoir) 
number densities, $\rhoc$ and $\rhop$ ($\rhop^r$), respectively.
As bulk thermodynamic parameters, we use the colloid packing fraction,
$\eta_C=(4\pi/3)\rhoc\Rc^3$, and effective polymer packing fractions,
$\eta_P=(4\pi/3)\rho_P (R_g^r)^3$ and $\eta_P^r=(4\pi/3)\rho_P^r (R_g^r)^3$,
where $R_g^r$ denotes the root-mean-square (rms) radius of gyration 
in the reservoir, a quantity that is directly accessible in experiments.  
The reservoir polymer-to-colloid size ratio, $R_g^r/\Rc$, 
provides a useful control parameter for tuning 
interparticle interactions, and thus thermodynamics.

\section{Theory}
\label{SECtheory}
\subsection{Density Functional Theory}
\label{SECdensityfunctional}
To investigate thermodynamic properties of the model system, 
we focus on the Helmholtz free energy as a functional of 
the inhomogeneous density profiles: the colloid density, 
$\rhoc(\rv)$, and a continuum of polymer densities, 
$\{\rhop(\rv;\Rp)\}$, indexed by $\Rp$ and normalized to 
the mean polymer density via
$V^{-1}\int{\rm d}^3 r \int_0^{\infty}{\rm d}\Rp\ \rhop(\rv;\Rp)=\rhop$.
It is convenient to separate the total free energy functional
into three terms:
\begin{eqnarray}
F[\rhoc(\rv),\{\rhop(\rv;\Rp)\}] &=& 
F_{\rm id}\left[\rhoc(\rv),\{\rhop(\rv;\Rp)\}\right] + 
F_{\rm ex}\left[\rhoc(\rv),\{\rhop(\rv;\Rp)\}\right]
\nonumber \\
&+& \int {\rm d}^3r \int_0^{\infty}{\rm d}\Rp\ \rhop(\rv;\Rp) 
\fch(\rv;\Rp;[\rhoc(\rv)]),
\label{Ftotal}
\end{eqnarray}
where $F_{\rm id}$ is the ideal free energy, $F_{\rm ex}$ is 
the excess free energy in the AO model generalized to polydisperse 
(but incompressible) polymer, and $\fch(\rv;\Rp;[\rhoc(\rv)])$ is 
the local free energy of a compressible polymer chain.
The third term in Eq.~(\ref{Ftotal}) stems from the {\em internal} 
degrees of freedom of the chains and is formally equivalent to 
the contribution of an external potential acting on the polymers.  
The ideal free energy, associated with (centre-of-mass) 
translational and mixing entropy of the colloids and polymers, 
is given exactly by
\begin{eqnarray}
 \beta F_{\rm id} &=& \int {\rm d}^3 r\ \rhoc(\rv)
\left[\ln\left(\rhoc(\rv)\Lambda_\colloid^3\right)-1\right]
\nonumber \\ 
&+& \int {\rm d}^3 r \int_0^{\infty}{\rm d}\Rp\ \rhop(\rv;\Rp)
\left[\ln\left(\rhop(\rv;\Rp)\Lambda_P^3\right)-1\right],
\end{eqnarray}
where $\beta\equiv 1/\kt$, $k_{\rm B}$ is Boltzmann's constant, 
and $\Lambda_C$ and $\Lambda_P$ are the respective 
colloid and polymer thermal wavelengths.

The excess free energy, arising from interactions, must be 
approximated.  For this purpose, we adopt a geometry-based DF 
approach, which is immediately applicable to multi-component systems.  
As a basis, we start from a recently-proposed DF theory~\cite{schmidt00cip}
for a binary CP mixture in the AO model, the homogeneous limit 
of which is equivalent to the free volume theory 
of Lekkerkerker \etal~\cite{lekkerkerker92}.
Here we generalize this theory~\cite{schmidt00cip} to 
mixtures of monodisperse colloids and polydisperse polymers.
Following previous work~\cite{rosenfeld89,schmidt00cip,RSLTshort,RSLTlong},
the excess free energy is expressed in the form
\begin{equation}
\beta F_{\rm ex} =
  \int {\rm d}^3 x \, \Phi \left( \{ \n{\fumindex}{\colloid}(\xv) \},
  \{ \n{\nu}{\polymer}(\xv) \}\right),
  \label{EQfexc}
\end{equation}
where the excess free energy density, $\Phi$, is a function of a 
set of weighted densities for colloid and polymer species.  
The weighted densities are defined as convolutions, with respect to 
geometric weight functions of the actual density profiles and, 
in the case of polymers, integration over radius of gyration:
\begin{eqnarray}
  \n{\fumindex}{\colloid}(\xv) &=&
  \int {\rm d}^3 r \, \rhoc(\rv) \, \w{\fumindex}{} (\xv-\rv;\Rc),\\
  \n{\fumindex}{\polymer}(\xv) &=& 
  \int {\rm d}^3 r \int_0^{\infty}{\rm d}\Rp\, \rhop(\rv;\Rp) \, 
\w{\fumindex}{} (\xv-\rv;\Rp).
\end{eqnarray}
We use standard fundamental-measure weight 
functions~\cite{rosenfeld89,tarazona00}, 
$\w{\fumindex}{}(\rv;R)$, $\nu=$0,1,2,3,v1,v2,m2, 
for spheres of radius $R$, given by
\begin{equation}
\w{2}{}(\rv;R) = \delta(R-r), \qquad
\w{3}{}(\rv;R) = \Theta(R-r), 
\end{equation}
\begin{equation}
\wv{2}{}(\rv;R)= \delta(R-r) \, \frac{\rv}{r},  \qquad
\wm{2}{}(\rv;R)= \delta(R-r) \left( \frac{\rv\rv}{r^2} - 
\frac{\mymat{1}}{3}
\right),
\end{equation}
where $r=|\rv|$, $\delta(r)$ is the Dirac distribution, 
$\Theta(r)$ is the step function, and $\mymat{1}$ is the identity matrix.  
Further, linearly dependent weight functions are 
$\w{1}{}(\rv;R) = \w{2}{}(\rv;R)/(4 \pi R)$, 
$\w{0}{}(\rv;R) = \w{1}{}(\rv;R)/R$, and
$\wv{1}{}(\rv;R) = \wv{2}{}(\rv;R)/(4 \pi R)$.
The weight functions have dimensions
$({\rm length})^{\nu-3}$ and differ in tensorial rank:
$\w{0}{}, \w{1}{}, \w{2}{}$, and $\w{3}{}$ are scalars; $\wv{1}{}$ 
and $\wv{2}{}$ are vectors; and $\wm{2}{}$ is a (traceless) matrix.

The excess free energy density in Eq.~(\ref{EQfexc})
separates naturally into three parts:
\begin{equation}
  \Phi = \Phi_1+\Phi_2+\Phi_3,
  \label{EQphi}
\end{equation}
which are defined as
\begin{eqnarray}
  \Phi_1 &=& \sum_{i= \colloid,\polymer} \n{0}{i} \,
  \myphi{i}\left(\n{3}{\colloid}, \n{3}{\polymer}\right), \label{EQphi1}
\\
  \Phi_2 &=& \sum_{i,j= \colloid,\polymer}
  \left( \n{1}{i} \n{2}{j} - \nv{1}{i} \cdot \nv{2}{j} \right) \,
  \myphi{ij}\left(\n{3}{\colloid}, \n{3}{\polymer}\right),
\\
\Phi_3 &=&\frac{1}{8\pi} \sum_{i,j,k= \colloid,\polymer}
          \left(
            \frac{1}{3}\n{2}{i}\n{2}{j}\n{2}{k} -
            \n{2}{i} \, \nv{2}{j} \cdot \nv{2}{k} 
\right. \nonumber \\ &+& \left.
          \frac{3}{2}\left[
            \nv{2}{i} \nmtl{2}{j} \nv{2}{k} -
            \tr \left( \nmtl{2}{i} \nmtl{2}{j} \nmtl{2}{k} \right)
          \right]
          \right) \,\myphi{ijk}\left(\n{3}{\colloid},
\n{3}{\polymer}\right).
\label{EQphi3}
\end{eqnarray}
Here $\tr$ denotes the trace operation and 
\begin{equation}
  \myphi{i\ldots k}(\etac, \etap) =
  \frac{\partial^m}
    {\partial \etai \cdots \partial \speciesindex{\eta}{k}}
    \beta F_{\rm 0d}(\etac, \etap)
  \label{EQmyphiDefinition}
\end{equation}
are derivatives of the zero-dimensional (0d) excess free energy,
\begin{equation}
    \beta F_{\rm 0d}(\etac, \etap) =
     (1-\etac-\etap)\ln(1-\etac) + \etac.
  \label{EQzerodf}
\end{equation}
Equations (\ref{EQfexc})-(\ref{EQzerodf}) completely specify 
the excess free energy in Eq.~(\ref{Ftotal}).

\subsection{Application to Bulk Fluids}
\label{SEChomogeneous}
For bulk fluid states, the density profiles are spatially constant.
In the homogeneous limit [$\rhoc(\rv)\to\rhoc$ and 
$\rhop(\rv;\Rp)\to\rhop(\Rp)$],
Eq.~(\ref{Ftotal}) yields the bulk fluid free energy density:
\begin{eqnarray}
\frac{\beta F}{V}~&=&~\rhoc\left[\ln\left(\rhoc\Lambda_C^3\right)-1\right] + 
\int_0^{\infty}{\rm d}\Rp\,\rhop(\Rp)
\left[\ln\left(\rhop(\Rp)\Lambda_P^3\right)-1\right]
\nonumber \\
 &+&~\phihs(\etac) - \int_0^{\infty}{\rm d}\Rp\, 
\rhop(\Rp)\left[\ln\alpha(\Rp;\etac)-\fch(\Rp;\etac)\right],
\label{FVhomogeneous1}
\end{eqnarray}
which is still a functional of the polymer density distribution, 
$\rhop(\Rp)$.  Here $\phihs(\etac)$ is the excess free energy density 
of the pure HS system, given by
\begin{equation}
  \phihs(\etac) = \frac{3 \ec 
[3 \ec (2-\ec) - 2(1-\ec)^2\ln(1-\ec)]}{8 \pi R_C^3 (1-\ec)^2},
\label{phihs}
\end{equation}
the same result as in the scaled-particle and Percus-Yevick 
compressibility approximations~\cite{HM},
and $\alpha(\Rp;\etac)$ is the free volume fraction, {\it i.e.}, 
the fraction of the total volume not excluded to the polymer
by the HS colloids, given implicitly by
\begin{equation}
\ln \alpha(\Rp;\etac) = \ln(1-\ec) - \sum_{m=1}^3 C_m \gamma^m,
\label{EQalpha}
\end{equation}
where $\gamma=\ec/(1-\ec)$ and the coefficients are polynomials
in the polymer-to-colloid size ratio, $q=\Rp/\Rc$: 
$C_1=3q+3q^2+q^3$, $C_2=9q^2/2+3q^3$, and $C_3=3q^3$.

It is worth noting that the free volume fraction is related to 
the polymer one-particle direct correlation function, 
\begin{equation}
c_P^{(1)}(\rv;\Rp;[\rhoc(\rv)])=-\beta \frac{\delta F_{\rm ex}}
{\delta \rhop(\rv;\Rp)},
\label{c1p}
\end{equation}
which has its physical origin in colloid-polymer correlations.
Substituting our DF approximation for $F_{\rm ex}$ into Eq.~(\ref{c1p})
and taking the homogeneous limit, we obtain
\begin{equation}
c_P^{(1)}(\Rp;\etac) = 
-\sum_{\fumindex} \frac{\partial \Phi}{\partial 
\n{\fumindex}{\polymer}} * \w{\fumindex}{}(\Rp)
= \ln\alpha(\Rp;\etac), 
\end{equation}
where $*$ denotes a convolution.  Because our approximate excess free energy 
functional is linear in the polymer density, $c_P^{(1)}(\Rp;\etac)$
is independent of $\rhop(\Rp)$, depending only on the colloid density.

In equilibrium, the polymer density distribution in Eq.~(\ref{FVhomogeneous1}) 
is fixed by the Euler-Lagrange equation for the polymers:
\begin{equation}
\frac{\delta (F/V)}{\delta \rhop(\Rp)} = \mu_P(\Rp),
\label{Euler1}
\end{equation}
where $\mu_P(\Rp)$ is the chemical potential of polymers with
radius of gyration $\Rp$.
Substituting Eq.~(\ref{FVhomogeneous1}) into Eq.~(\ref{Euler1}),
we have
\begin{equation}
\ln\left(\rhop(\Rp;\etac)\Lambda_P^3\right) - \ln\alpha(\Rp;\etac) + 
\beta\fch(\Rp) = \beta\mu_P(\Rp),
\label{Euler2}
\end{equation}
where $\rhop(\Rp;\etac)$ is the {\em equilibrium} polymer density 
distribution at bulk colloid volume fraction $\etac$.
Here we make the simplifying assumption that the chain free energy of 
a polymer with given $\Rp$ in the system is the same as that of 
an equal-sized polymer in the reservoir,
and thus independent of colloid concentration.
Because the system and reservoir must be in equilibrium 
with respect to polymer exchange, the right side of 
Eq.~(\ref{Euler2}) can be equated with the chemical potential 
of the corresponding polymer species in the reservoir:
\begin{equation}
\mu_P(\Rp) = \mu_P^r(\Rp) =
\beta^{-1}\ln\left(\rhop^r(\Rp)\Lambda_P^3\right) + \fch(\Rp),
\label{EQmup}
\end{equation}
where $\rhop^r(\Rp)$ denotes the polymer density distribution 
in the reservoir.
Further progress is facilitated by factoring the 
reservoir polymer density distribution, according to 
\begin{equation}
\rhop^r(\Rp) = \rhop^r P^r(\Rp),
\label{EQrhop0}
\end{equation}
where $\rhop^r$ and $P^r(\Rp)$ are the mean density and
radius-of-gyration probability distribution, respectively, 
of polymers in the reservoir.
The advantage of Eq.~(\ref{EQrhop0}) is that the probability distribution 
is simply related to the chain free energy via~\cite{doi-edwards,yamakawa}
\begin{equation}
\beta\fch(\Rp)=-\ln P^r(\Rp) + {\rm constant~independent~of}~\Rp,
\label{fchain}
\end{equation}
the arbitrary additive constant being chosen to normalize the 
distribution: $\int_0^\infty {\rm d}\Rp\, P^r(\Rp)=1$.
Combining Eqs.~(\ref{Euler2}) and (\ref{EQmup}), and using
Eqs.~(\ref{EQrhop0}) and (\ref{fchain}), we obtain
\begin{equation}
\rhop(\Rp;\etac) = \rhop^r \, \alpha(\Rp;\etac) P^r(\Rp)
= \alpha(\Rp;\etac) \rhop^r(\Rp). 
\label{EQrhop}
\end{equation}
The mean polymer densities in the system and reservoir are 
now seen to be related via
\begin{equation}
\rhop(\etac) = \int_0^{\infty}{\rm d}\Rp\, \rhop(\Rp;\etac)
= \rhop^r \, \alpha_{\rm eff}(\etac),
\label{rhop}
\end{equation}
where
\begin{equation}
\alpha_{\rm eff}(\etac) = \int_0^{\infty}{\rm d}\Rp\, 
\alpha(\Rp;\etac) P^r(\Rp)
\label{alphaeff}
\end{equation}
is an {\em effective} free volume fraction. 
Equation~(\ref{rhop}) is a generalization of the 
incompressible-polymer relation, $\rhop(\etac)=\rhop^r\alpha(\etac)$,
from free volume theory~\cite{lekkerkerker92,schmidt00cip}.
Factoring $\rhop(\Rp;\etac)$ according to
\begin{equation}
\rhop(\Rp;\etac) = \rhop(\etac) P(\Rp;\etac),
\end{equation}
serves to define 
\begin{equation}
P(\Rp;\etac) = \frac{\alpha(\Rp;\etac)}{\alpha_{\rm eff}(\etac)} \, P^r(\Rp)
\label{PRp}
\end{equation}
as the normalized radius-of-gyration probability distribution 
of polymers in the system.
Equation~(\ref{PRp}) makes manifest that the radius-of-gyration 
distribution in the system depends on colloid concentration and
differs from the distribution in the reservoir.

Finally, substituting Eqs.~(\ref{fchain}) and (\ref{EQrhop}) into
Eq.~(\ref{FVhomogeneous1}) and rearranging, we obtain
the equilibrium bulk fluid free energy density,
\begin{equation}
\frac{\beta F}{V} = \rhoc\left[\ln\left(\rhoc\Lambda_C^3\right)-1\right] + 
\rhop^r\alpha_{\rm eff}(\etac)\left[\ln(\rhop^r\Lambda_P^3)-1\right] 
+\phihs(\etac).
\label{FVhomogeneous2}
\end{equation}
Equations~(\ref{PRp}) and (\ref{FVhomogeneous2}) are the main 
theoretical results of the paper and our basis for calculating 
polymer size distributions and phase behaviour in Sec.~\ref{SECresults}.
Practical implementation, however, first requires specification 
of $P^r(\Rp)$.

\subsection{Radius of Gyration Probability Distribution}
\label{SECchain}
To approximate the radius-of-gyration probability distribution, 
we appeal to the statistical mechanics of a freely-jointed chain.
The configuration of a chain of $N$ links, each link of length $a$, 
can be specified by a set of position vectors, 
$\{\Rv_i\}=(\Rv_0\ldots\Rv_N)$, of the joints.
For a given configuration, the radius of gyration, $\Rp$, is 
defined by~\cite{doi-edwards} 
\begin{equation}
 \Rp^2 = \sum_{i=1}^N (\Rv_i-\Rv_{\rm cm})^2,
\end{equation}
where 
$\Rv_{\rm cm}= N^{-1} \sum_{i=1}^N \Rv_i$ is the 
centre-of-mass position of the chain. 
Averaged over configurations, the radius of gyration 
follows the probability distribution, $P^r(\Rp)$. 
Moments of the distribution are defined as
\begin{equation}
 \langle\Rp^n\rangle  = \int_0^{\infty}{\rm d}\Rp\, \Rp^n\, P^r(\Rp),
\end{equation}
the second ($n=2$) moment being related to the rms radius of gyration
of polymers in the reservoir~\cite{doi-edwards,yamakawa},
$\Rg^r = \sqrt{\langle\Rp^2\rangle} = a\sqrt{N/6}$.
For polymers in the system, the rms radius of gyration is defined as
\begin{eqnarray}
\Rg(\etac)  &=& \left(\int_0^{\infty}{\rm d}\Rp\, 
\Rp^2\, P(\Rp;\etac)\right)^{1/2} \nonumber \\
&=&  \left(\frac{1}{\alpha_{\rm eff}(\etac)}
\int_0^{\infty}{\rm d}\Rp\, \Rp^2\, \alpha(\Rp;\etac) \, 
P^r(\Rp)\right)^{1/2},
\end{eqnarray}
which clearly varies with bulk colloid concentration.
Note that $\Rg(\etac=0)=\Rg^r$.

In general, the radius of gyration is a more appropriate measure 
of polymer size than the end-to-end displacement, 
which is not well defined for branched polymers.
However, in contrast to the simple Gaussian distribution of 
the end-to-end displacement, the radius-of-gyration distribution 
is nontrivial, even for ideal chains.  
Flory and Fisk~\cite{flory-fisk66,flory} first proposed an empirical 
approximation for $P^r(\Rp)$ based on the exact even moments 
calculated by Fixman~\cite{fixman62}.
Subsequently, Fujita and Norisuye~\cite{yamakawa,fujita70}
calculated the distribution {\it exactly}, obtaining the
analytical result 
\begin{eqnarray}
P^r(\Rp)~&=&~\frac{1}{\sqrt{2}\,\pi\Rg^r\,t^3}
\sum_{k=0}^{\infty}\frac{(2k+1)!}{(2^kk!)^2}(4k+3)^{7/2}\exp(-t_k) 
\nonumber \\
 &\times&~\left[\left(1-\frac{5}{8t_k}\right)K_{1/4}(t_k)+
\left(1-\frac{3}{8t_k}\right)K_{3/4}(t_k)\right],
\label{Fujita}
\end{eqnarray}
where $t=(\Rp/\Rg^r)^2$, $K_n$ are the modified Bessel functions 
of the second kind, and $t_k=(4k+3)^2/(8t)$.
To confirm Eq.~(\ref{Fujita}), we have carried out 
Monte Carlo simulations of an ideal chain for $N=100$ and 1000.
Histograms of $\Rp$, generated from $10^6$ independent configurations,
are in essentially perfect agreement with 
the analytical expression~\cite{schmidt-denton02}. 

Equation (\ref{Fujita}) specifies the radius-of-gyration probability
distribution and thus, together with Eq.~(\ref{fchain}), 
the chain free energy.
We emphasize that in modelling the chains as freely jointed, and so 
applying the ideal-chain distribution [Eq.~(\ref{Fujita})] to polymers 
mixed with colloids, we implicitly neglect any effect of colloidal
confinement on the basic shape of the distribution.  In particular,
confinement-induced nonspherical distributions are neglected,
although variation in the polymer size is allowed.

\section{Results: Demixing Phase Behaviour and Polymer Size Distribution}
\label{SECresults}
The general conditions for phase coexistence are 
equality of the total pressure, 
\begin{equation}
p_{\rm tot}=-\frac{F}{V}+\sum_{i={C,P}} \rho_i
~\frac{\partial (F/V)}{\partial \rho_i}
\label{ptot}
\end{equation}
and of the chemical potentials for each species,
\begin{equation}
\mu_i=\frac{\partial (F/V)}{\partial \rho_i}, \quad i=C,P,
\label{mui}
\end{equation}
between the coexisting phases.
Equilibrium between phases I and II requires
$p_{\rm tot}^{\rm I} = p_{\rm tot}^{\rm II}$ and $\mu_C^{\rm I} =
\mu_C^{\rm II}$, equality of the polymer chemical potentials 
being enforced by Eq.~(\ref{Euler1}).

In practice, bulk fluid phase diagrams can be computed as follows.
The effective free volume fraction, $\alpha_{\rm eff}(\etac)$, 
is first determined by substituting Eqs.~(\ref{EQalpha}) 
and (\ref{Fujita}) into Eq.~(\ref{alphaeff}). 
Then, for a given reservoir polymer density, $\rhop^r$, 
Eqs.~(\ref{FVhomogeneous2}), (\ref{ptot}), and (\ref{mui})
($i=C$) are solved numerically for the coexisting colloid packing fractions,
$\etac^{\rm I}$ and $\etac^{\rm II}$.
Finally, Eq.~(\ref{rhop}) converts from reservoir to system representation, 
giving the corresponding system polymer packing fractions, 
$\etap^{\rm I}$ and $\etap^{\rm II}$.

\begin{figure}
  \begin{center}
    \includegraphics[width=0.5\columnwidth,angle=-90]{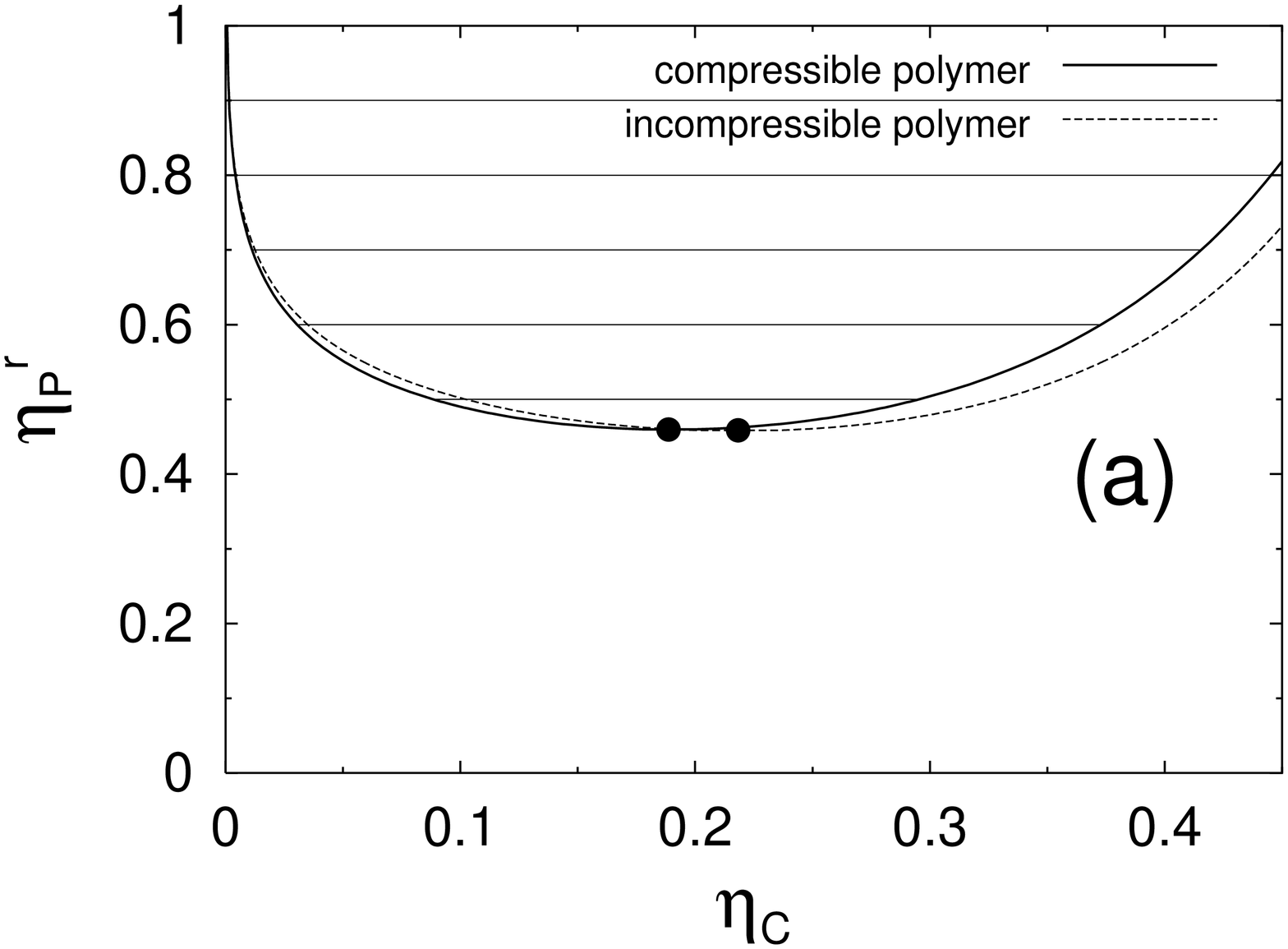}
    \includegraphics[width=0.5\columnwidth,angle=-90]{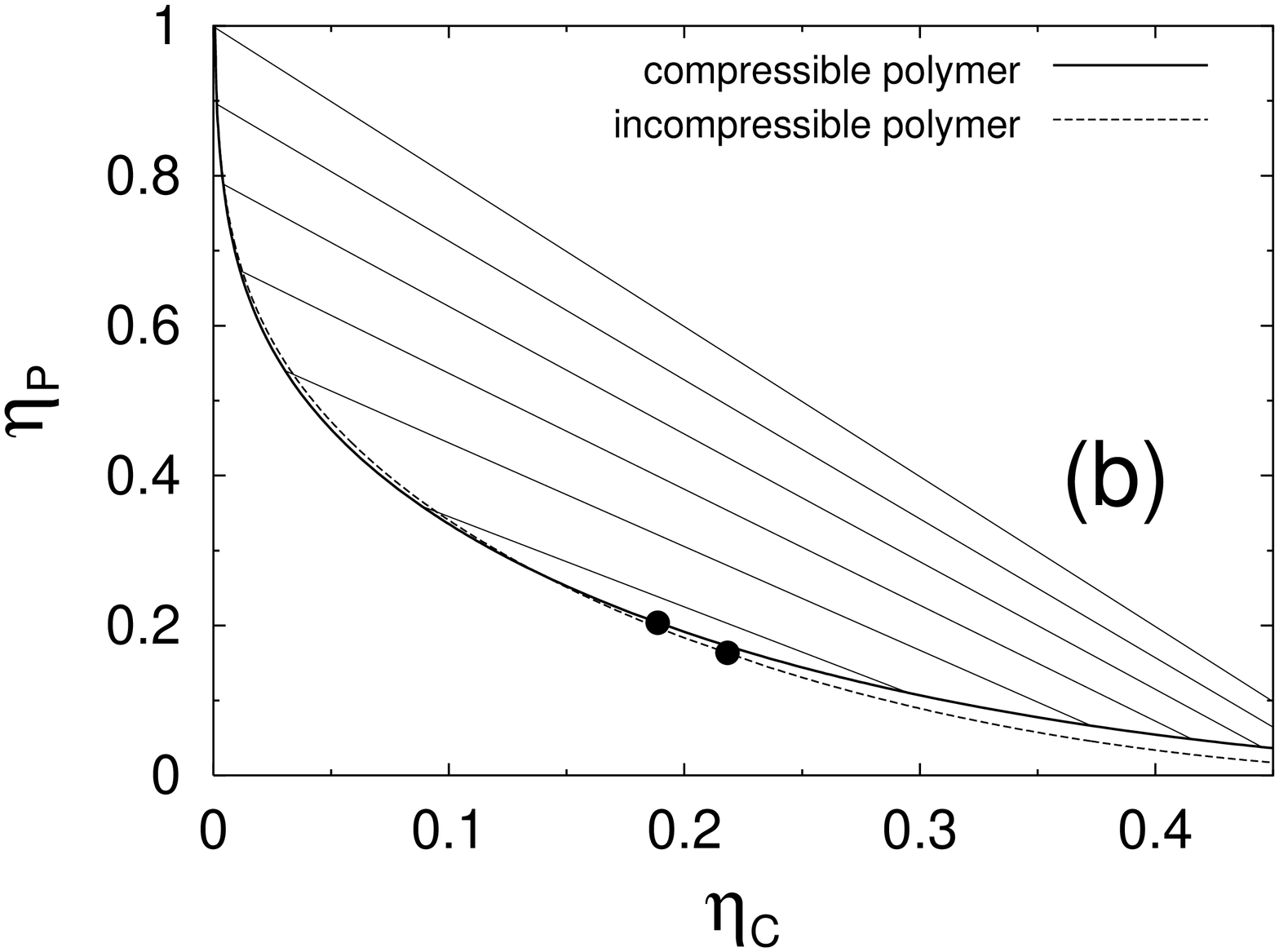}
    \caption{Fluid-fluid demixing phase diagram for rms reservoir
    polymer-to-colloid size ratio $R_g^r/R_C=0.5$. Solid curve: 
    binodal from the present theory, taking into account polymer
    compressibility; dashed curve: binodal from free volume theory 
    for the AO model with incompressible polymer. 
    Thin straight lines indicate tie lines between 
    coexisting phases; dots indicate critical points. 
    (a) Reservoir representation: 
    polymer reservoir packing fraction $\eta_P^r$ vs. 
    colloid packing fraction $\eta_C$; (b) system representation: 
    system polymer packing fraction $\eta_P$ vs. $\eta_C$.}
    \label{FIGpd05} \end{center}
\end{figure}

\begin{figure}
  \begin{center}
    \includegraphics[width=0.5\columnwidth,angle=-90]{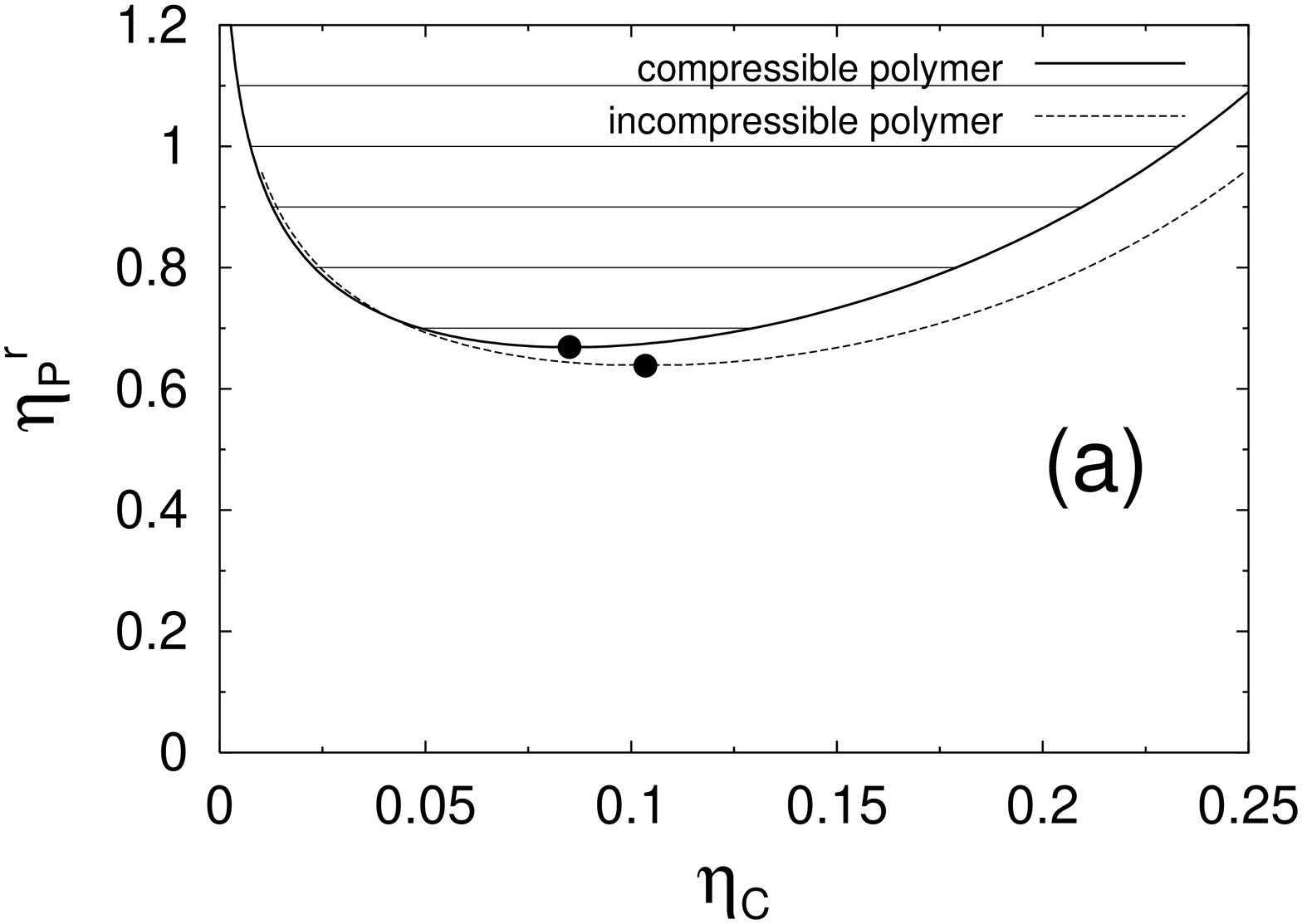}
    \includegraphics[width=0.5\columnwidth,angle=-90]{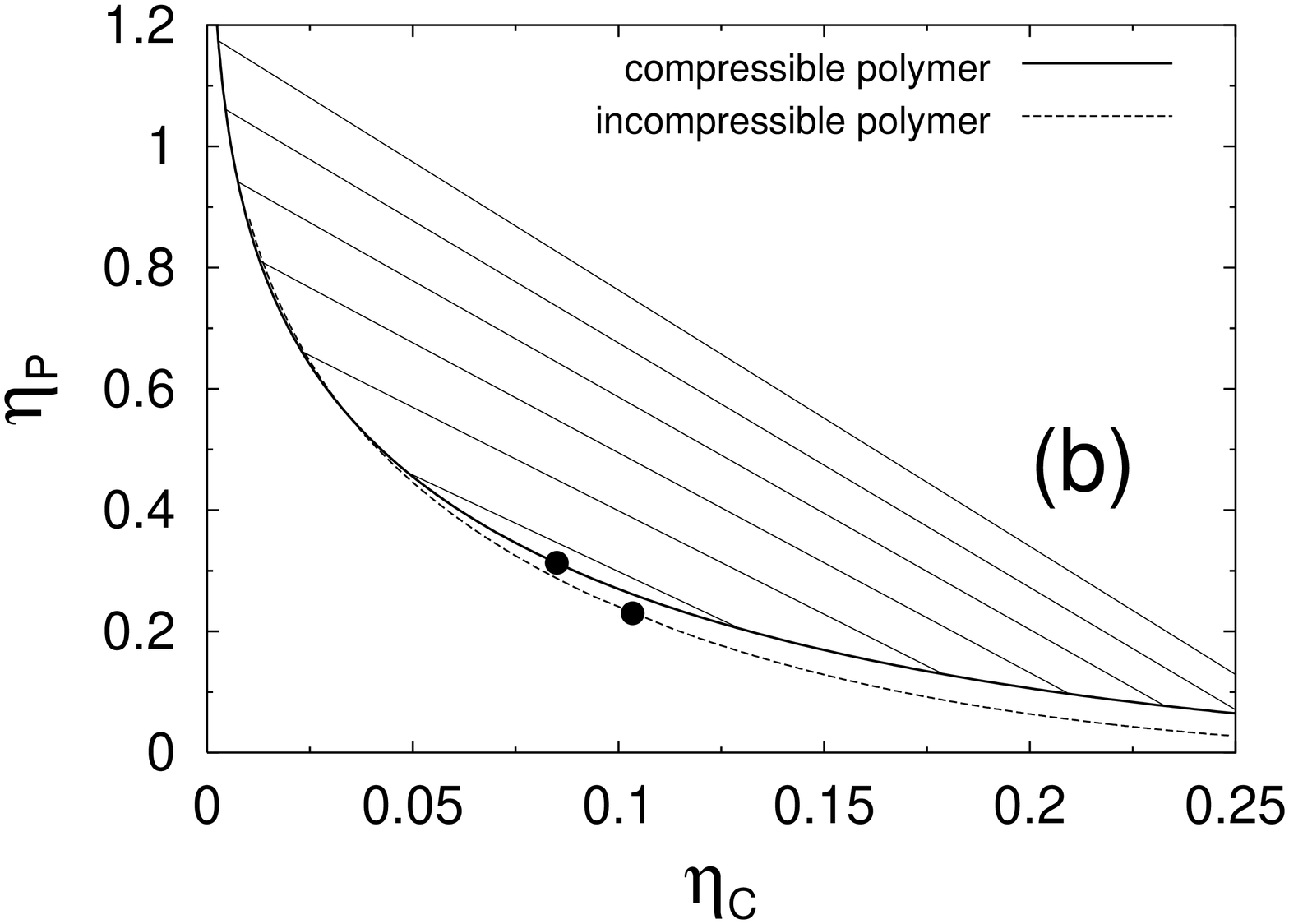}
    \caption{Same as Fig.~\protect\ref{FIGpd05}, but for $R_g^r/R_C=1$.
    Note the change in scale compared with Fig.~\ref{FIGpd05}.}
    \label{FIGpd1} \end{center}
\end{figure}

Figures~\ref{FIGpd05} and \ref{FIGpd1} present bulk fluid phase diagrams 
for rms reservoir polymer-to-colloid size ratios $R_g^r/R_C=$0.5 and 1, 
in both the system and reservoir representations.
For comparison, we include demixing binodals both from the present theory, 
which takes into account polymer compressibility, and from the
free volume theory of the AO model with incompressible polymer.
For sufficiently high colloid and polymer packing fractions,
the system demixes into colloid-rich (liquid) and colloid-poor (vapour)
phases.  We disregard here the liquid-solid branch of the phase diagram, 
since our size ratios are sufficiently high that fluid-fluid 
demixing can be assumed stable.
The effect of polymer compressibility evidently is to shift
the colloidal liquid branch of the binodal toward higher polymer density,
stabilizing the system against demixing.  
Interpreted in terms of effective depletion-induced attraction
between pairs of colloids, polymer compression shortens the range
of attraction, tending to favour mixing. 
In passing, it may be anticipated that additional degrees of freedom
allowing for nonspherical polymer conformations would tend to 
further stabilize the mixture if, as an alternative to bulk demixing,
the polymer may simply distort its shape.

\begin{figure}
  \begin{center}
    \includegraphics[width=0.5\columnwidth,angle=-90]{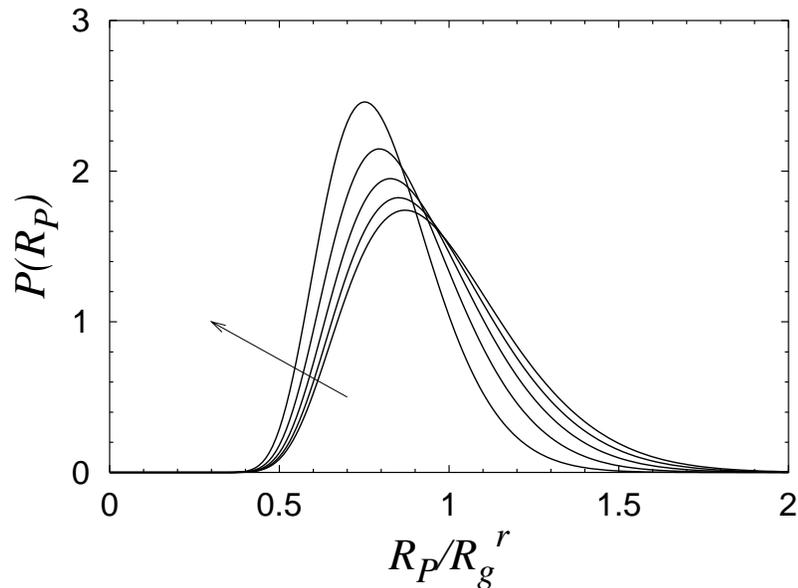}
    \caption{Normalized probability distribution, $P(R_P)$, of scaled polymer 
    radius of gyration, $R_P/R_g^r$ for rms reservoir polymer-to-colloid 
    size ratio $R_g^r/R_C=0.5$ (corresponding to Fig.~\ref{FIGpd05}).
    Colloid packing fraction, $\eta_C$, increases in direction of arrow: 
    $\eta_C=$0, 0.1, 0.2, 0.3, and 0.4.  Corresponding rms polymer size ratios 
    decrease: $R_g/R_g^r=$1, 0.974, 0.938, 0.891, and 0.829.}
    \label{FIGpofr05} \end{center}
\end{figure}

\begin{figure}
  \begin{center}
    \includegraphics[width=0.5\columnwidth,angle=-90]{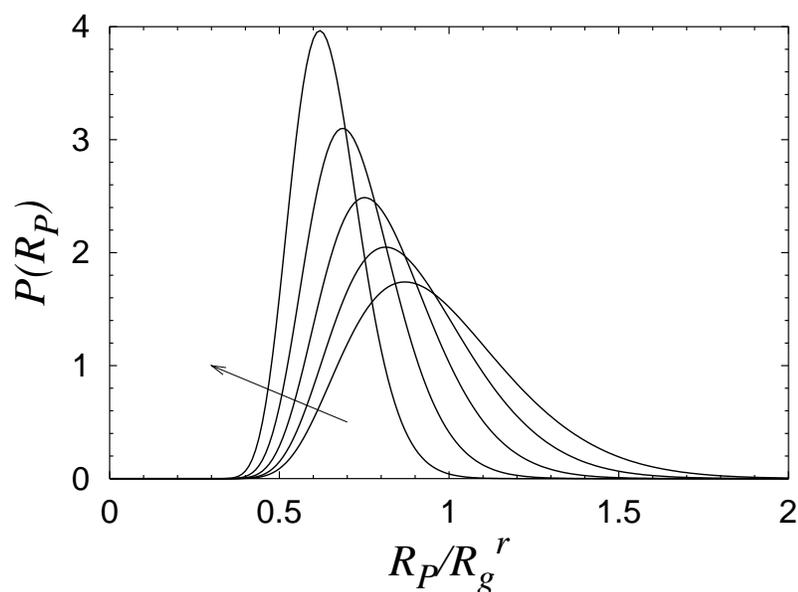}
    \caption{Same as Fig.~\ref{FIGpofr05}, but for $R_g^r/R_C=1$
    (corresponding to Fig.~\ref{FIGpd1}).  In direction of arrow,
    rms polymer size ratios are $R_g/R_g^r=$1, 0.915, 0.826, 0.738, and 0.653.
    Note the change in scale compared with Fig.~\ref{FIGpofr05}.}
    \label{FIGpofr1} \end{center}
\end{figure}

Figures~\ref{FIGpofr05} and \ref{FIGpofr1} show the corresponding
normalized polymer size distributions as a function of radius-of-gyration, 
scaled to the rms reservoir value.  For given rms reservoir size ratio,
with increasing colloid concentration, 
the distributions narrow and shift toward smaller radii, reflecting 
compression of the polymer by the confining colloids.
In the case of $R_g^r/R_C=0.5$, the polymer is compressed, relative to
its rms size in the reservoir, to $R_g/R_g^r=$0.974, 0.938, 0.891, and 0.829
at colloid packing fractions $\eta_C=$0.1, 0.2, 0.3, and 0.4, respectively.
For $R_g^r/R_C=1$, the polymer is even more strongly compressed to
$R_g/R_g^r=$0.915, 0.826, 0.738, and 0.653 at the same respective 
values of $\etac$.
Note that, for given $\etac$, $P(\Rp)$ does not depend on $\etap$ 
(nor on $\etap^r$).  Hence in the phase diagram, the polymer size 
distribution is constant along vertical lines (in both system and 
reservoir representations).  What changes, of course, is the overall 
prefactor in $\rhop(\Rp)$.  This invariance is an approximate feature 
of the theory resulting from the linearity of the excess free energy 
in polymer density.  In reality, the presence of the polymers 
may alter the confining structure of the colloids, in turn
modifying $P(\Rp)$.

\section{Conclusions}
\label{SECconclusions}
We have investigated the bulk fluid demixing behaviour of a model 
mixture of hard colloidal spheres and non-adsorbing polymer chains 
in a theta solvent.  The polymer component is modelled, on a 
mesoscopic level, as a polydisperse mixture of effective spheres 
with radii free to adjust to allow for colloid-induced compression.
The model goes beyond previous studies by treating the polymer 
radius of gyration as an internal degree of freedom that is 
polydisperse and varies between coexisting phases.
Like all effective-sphere models, however, our approach ignores 
details of the polymers on the segment level and thereby neglects
any effects of polymer shape anisotropy.  

To describe the model system, we have developed a geometry-based 
density-functional theory that incorporates polymer configurational 
free energy from consideration of the statistical mechanics of 
an ideal (random-walk) chain.
For simplicity, the polymer chain free energy is assumed to be 
insensitive to colloidal confinement.
Minimization of an approximate free energy functional with respect to 
the polymer size distribution yields both the equilibrium size distribution
-- modified by colloid-polymer interactions -- and the free energy, 
from which we compute bulk fluid phase diagrams.
Polymer compression is found to increase with increasing colloid 
concentration and polymer-to-colloid size ratio.
Correspondingly, the demixing fluid binodal shifts to higher 
polymer densities on the colloid-rich branch, favouring mixing. 

Predictions of the theory may be tested by experiment and simulation.
Although the predicted shift of the demixing binodal due to polymer 
compression would appear to improve agreement with limited experimental 
data near the theta point~\cite{ilett95}, more detailed measurements 
are certainly desirable.
Polymer size distributions could be probed using either 
light scattering, by index matching the colloids (but not polymers) 
to the solvent, or neutron scattering by selectively deuterating 
the polymer (hydrocarbon) backbones.
Detailed molecular simulations of segmented polymer chains mixed with 
hard spheres, along the lines of Refs.~\cite{meijer-frenkel91} and
\cite{dickman-yethiraj94}, could provide the cleanest tests of
our approximations and predictions.

Future applications to inhomogeneous colloid-polymer mixtures 
under the influence of external potentials (due to walls, gravity, etc.) 
and to liquid-solid transitions would be feasible and worthwhile.
Extensions of the theory could explore the sensitivity of 
the polymer chain free energy to colloidal confinement, polydispersity 
in chain length, influences of anisotropic polymer conformations, 
distinctions between linear and branched polymers
(as in colloid-star polymer mixtures~\cite{poon01,likos01,likos02}), 
and effects of polymer nonideality in good and poor solvents.

\ack
A helpful discussion with Dr Michael Rubinstein is gratefully acknowledged.

\newpage

\newpage

\newpage

\end{document}